\documentclass[twocolumn,showpacs,reprint,amsmath,amssymb,
 aps,preprintnumbers,amsmath,amssymb]{revtex4}
\usepackage{graphicx}
\usepackage{dcolumn}
\usepackage{bm}
\usepackage{sidecap}

\begin{document}

\preprint{APS/123-QED}

\title{In-beam spectroscopic studies of $^{44}$S nucleus \\}

\author{L. C\'aceres$^1$, D. Sohler$^2$, S.~Gr\'evy$^1$, O.~Sorlin$^{3,1}$, Zs.~Dombr\'adi$^2$, B.~Bastin$^{4,1}$, N.~L.~Achouri$^4$, J.~C.~Ang\'elique$^4$, F.~Azaiez$^3$, D.~Baiborodin$^5$, R.~Borcea$^6$, C.~Bourgeois$^3$, A.~Buta$^6$, A.~B\"urger$^{7,8}$, R.~Chapman$^{9}$, J.~C.~Dalouzy$^1$, Z.~Dlouhy$^5$, A. Drouard$^7$, Z.~Elekes$^2$, S.~Franchoo$^3$, L.~Gaudefroy$^{10}$, S.~Iacob$^6$, B.~Laurent$^{10}$, M.~Lazar$^6$, X.~Liang$^{9}$, E.~Li\'enard$^4$, J.~Mrazek$^5$, L. Nalpas$^7$, F.~Negoita$^6$, F. Nowacki$^{11}$, N.~A.~Orr$^4$, Y.~Penionzhkevich$^{12}$, Zs.~Podoly\'ak$^{13}$, F.~Pougheon$^3$, A.~Poves$^{14}$, P.~Roussel-Chomaz$^1$, M.~G.~Saint-Laurent$^1$, M.~Stanoiu$^{1,6}$, I.~Stefan$^{1,3}$}

\affiliation{$^1$Grand Acc\'el\'erateur National d'Ions Lourds (GANIL), CEA/DSM-CNRS/IN2P3, Caen, France}
\affiliation{$^2$Institute of Nuclear Research, H-4001 Debrecen, Pf.51, Hungary}
\affiliation{$^3$Institut de Physique Nucl\'eaire, IN2P3-CNRS, F-91406 Orsay Cedex, France}
\affiliation{$^4$Laboratoire de Physique Corpusculaire, 6,bd du Mal Juin, F-14050 Caen Cedex, France}
\affiliation{$^5$Nuclear Physics Institute, AS CR, CZ-25068 Rex, Czech Republic}
\affiliation{$^6$Institute of Physics and Nuclear Engineering IFIN-HH, P.O. Box MG-6, 077125 Bucharest-Magurele,
Romania}
\affiliation{$^7$CEA Saclay, DAPNIA/SPhN, F-91191 Gif-sur-Yvette Cedex, France}
\affiliation{$^8$Helmholtz-Institut f\"ur Strahlen- und Kernphysik, Universit\"at Bonn, Nu$\beta$allee 14-16,
 D-53115 Bonn, Germany}
\affiliation{$^{9}$SUPA, School of Engineering, University of the West Scotland, Paisley PA1 2BE, Scotland, 
United Kingdom}
\affiliation{$^{10}$CEA, DAM, DIF, F-91297 Arpajon, France}
\affiliation{$^{11}$IPHC, BP28, F-67037 Strasbourg Cedex, France}
\affiliation{$^{12}$FLNR, JINR, 141980 Dubna, Moscow Region, Russia}
\affiliation{$^{13}$University of Surrey, GU2 7XH Guildford, United Kingdom}
\affiliation{$^{14}$Departamento de F\'isica Te\'orica, Universidad Aut\'onoma de Madrid, E-28049 Madrid, Spain}

\date{\today}
\begin{abstract}
The structure of the $^{44}$S nucleus has been studied at GANIL through the one proton knock-out reaction from a $^{45}$Cl secondary beam at 42~A$\cdot$MeV. The $\gamma$ rays following the de-excitation of $^{44}$S were detected in flight using the 70 BaF${_2}$ detectors of the Ch\^{a}teau de Cristal array.
An exhaustive $\gamma\gamma$-coincidence analysis allowed an unambiguous construction of the level scheme up to an excitation energy of 3301~keV. The existence of the spherical 2$^+_2$ state is confirmed and three new $\gamma$-ray transitions connecting the prolate deformed 2$^+_1$ level were observed. Comparison of
the experimental results to shell model calculations further supports a prolate and spherical shape
coexistence with a large mixing of states built on the ground state band in $^{44}$S.

\end{abstract}


\maketitle

\section{Introduction}
In the 50's, Goeppert-Mayer~\cite{mayer} and Haxel, Jensen, 
and Suess~\cite{suess} were able to obtain a description of the nuclear magic
numbers by adding a strong spin-orbit interaction to a central mean field potential. This leads to the
creation of the so-called spin-orbit gaps with magic numbers 2, 8, 20, 28, 50, 82 and 126. Stable nuclei with magic number of protons and/or neutrons were found to 
be spherical, having a closed shell configuration. This simple view of indestructible magic numbers has been modified over the last 30 years due to new spectroscopic information on exotic nuclei such as $^{32}_{12}$Mg$_{20}$~\cite{32Mg,32MgB}. A reduction of the spherical $N=20$ gap was invoked to explain the deformation of $^{32}_{12}$Mg where a gain in correlation energy, due to quadrupole excitations obtained when promoting nucleons across the shell gap, is larger than the loss in energy required to cross that shell gap. Therefore, if the spherical shell gap is reduced, and/or the two orbits forming the gap are separated by two units of angular momentum, as for all spin-orbit magic numbers, the nuclear deformation is favored. These two conditions are fulfilled also in $^{42}$Si, which was found to be deformed~\cite{beyhan, bazin}. One of the key nuclei to understand how deformation sets in for the N~=~28 isotones between the spherical $^{48}$Ca and $^{42}$Si is $^{44}$S. 
This nucleus has been investigated using different experimental
 approaches, such as $\beta$-decay~\cite{sorlin,grevy1}, Coulomb excitation~\cite{glas}, in-beam $\gamma$-ray spectroscopy~\cite{dora}, isomer studies~\cite{grevy,force}, electron spectroscopy~\cite{force} and the two proton knock-out reaction~\cite{santiago}. The works of Ref.~\cite{glas, dora} showed that $^{44}$S has a deformed ground state. 
The study of the decay of the 0$^+_2$ level~\cite{force} and the B(E2:~2$_1^+\rightarrow$0$^+_1$) reduced transition probability~\cite{glas} led to the conclusion, by means of shell model calculations, that $^{44}$S exhibits a prolate ground state and a 0$^+_2$ spherical isomeric level.
 In addition, shell-model calculations predicted the existence of a deformed band built on top of the ground state
 and a 2$^+_2$ level of spherical character above the 0$^+_2$ state. A candidate for the 2$^+_2$ level has been proposed in Ref.~\cite{santiago}. The aim of the present work is to confirm the existence of this state, among others found in Ref.~\cite{santiago}, and to discover new-high energy levels connecting the deformed and spherical bands in $^{44}$S. All this information is relevant to further understand shape~-~coexistence in this nucleus. The authors of Ref.~\cite{santiago} used a two proton knock-out reaction from a secondary beam  of $^{46}$Ar to populate excited states in $^{44}$S while, in the present work, a one proton knock-out reaction $^{45}$Cl(-1p)$^{44}$S was employed. This different reaction mechanism is expected to populate additional states in $^{44}$S leading to a complementary understanding of the structure of this nucleus.

\section{Experimental method}

\begin{figure}
\includegraphics[width=0.50\textwidth]{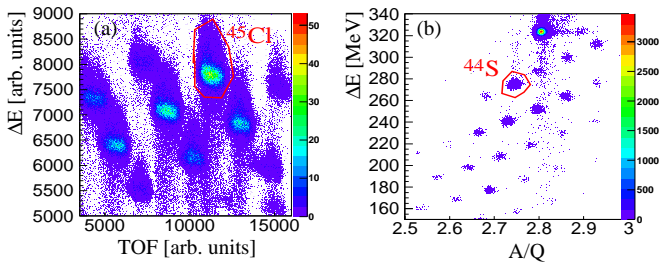}
\caption{\label{fig:z_aoq}(Color online) (a): Energy loss ($\Delta$E) versus time of flight (TOF) particle identification plot of the ions produced in the fragmentation of $^{48}$Ca beam on C target and separated in the ALPHA spectrometer. (b): Energy loss ($\Delta$E) versus A/Q plot of the ions produced in the fragmentation of the $^{45}$Cl secondary beam in the Be secondary target and separated in the SPEG spectrometer.}
\end{figure}

The experiment was performed at the Grand Acc\'el\'erateur National d'Ions Lourds (GANIL, France) facility.
A $^{48}$Ca$^{19+}$ primary beam at 60~MeV/u and 3.8~$\mu$A average intensity impinged on a 200~mg/cm$^2$ C target located between the two superconductor solenoids of the SISSI~\cite{sissi} device.
The reaction products were separated by means of the B$\rho$-$\Delta$E-B$\rho$ method~\cite{dufour, Anne} in the ALPHA
 spectrometer. The ion identification was performed on an event-by-event basis by measuring their time-of-flight (TOF) over an 80~m flight path using two micro-channel plates and energy loss ($\Delta$E) in a 50~$\mu$m Si detector located at the entrance of the SPEG spectrometer. An example of the particle identification plot for the reaction residues is shown in Fig.~\ref{fig:z_aoq}~(a). 
Afterwards, the selected fragments underwent secondary reactions in a 195~mg/cm$^2$ $^9$Be target placed at the Si detector position. The $\Delta$E and positions of the secondary ions were measured at the final focal plane of SPEG in ionization and drift chambers, respectively. A plastic scintillator was used to determine the TOF and residual energy. All of these measurements permitted a full reconstruction of the mass-to-charge ratio of the produced secondary ions (Fig.~\ref{fig:z_aoq}~(b)). 
Although, the magnetic rigidity of the spectrometer was optimized 
for $^{42}$Si transmission, the large acceptance of SPEG ($\Delta$p/p$\sim$7\%) spectrometer permitted 50\% of the $^{44}$S momentum distribution from the one proton knock-out reaction from $^{45}$Cl to be transmitted. This reaction is of particular interest as it provides spin and parity selectivity of the final states populated.
The inclusive $\sigma_{1p}$($^{45}$Cl$\rightarrow^{44}$S) cross section was measured to
 be 13(3)~mb. The prompt-$\gamma$ rays detected in coincidence with the nuclei identified at the final focal plane of SPEG
 were measured in 70~BaF$_2$ detectors of the Ch\^{a}teau de Cristal array positioned above and below 
the secondary target at an average distance of 25~cm. The efficiency of the
 array was determined using standard sources of $^{60}$Co and $^{152}$Eu to be 24\% at 1.33~MeV with an energy resolution of 15\% at 800~keV. A time resolution of 800~ps was obtained. 
\begin{figure}
\includegraphics[width=0.5\textwidth]{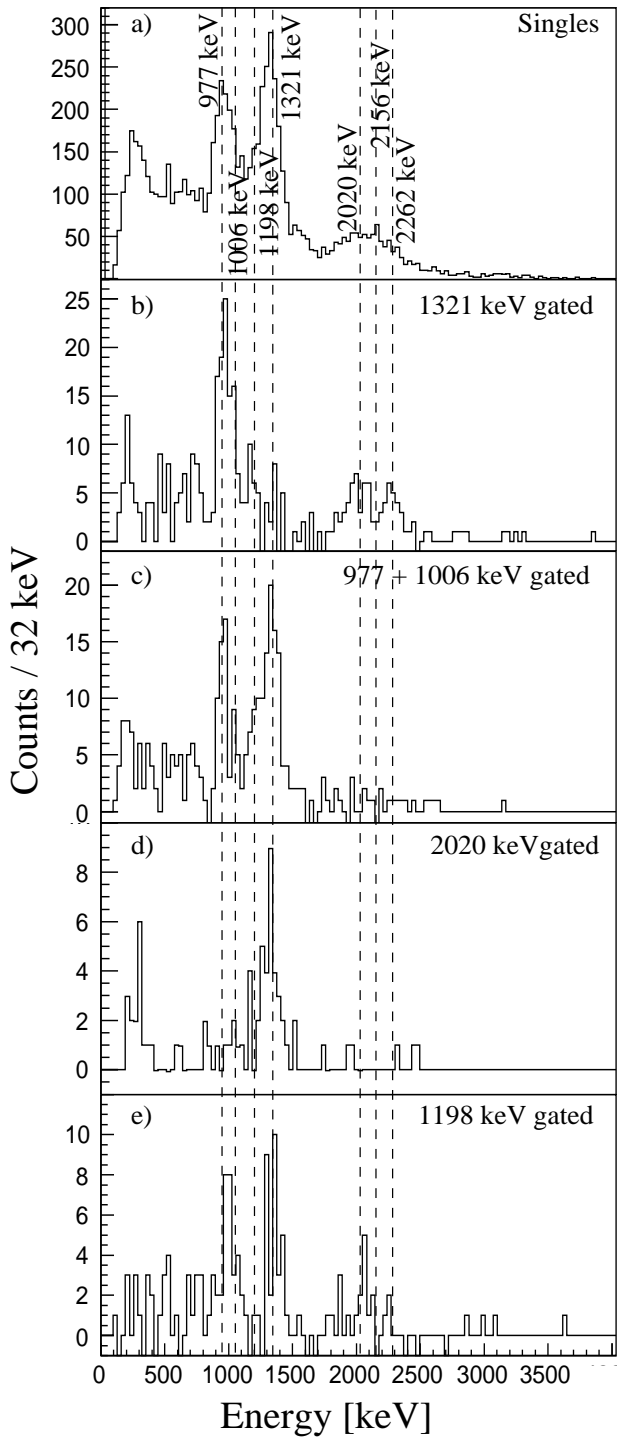}
\caption{\label{fig:energy} (a): Singles $\gamma$-ray spectrum in coincidence with $^{44}$S ions.
 Coincidence spectra gated by the (b) 1321, (c) summed 977 and 1006, (d) 1979 and (e) 1198~keV transitions, respectively.}
\end{figure}

\section{results}

The $\gamma$-ray singles spectrum in coincidence with $^{44}$S reaction products is shown in Fig.~\ref{fig:energy}(a). Transitions previously reported in Ref.~\cite{glas,dora} are confirmed with the exception of the $\gamma$-ray at 2632~keV that is not visible in the present work. However, a broad $\gamma$-ray distribution at high excitation energy is observed.
The construction of the level scheme shown in Fig.~\ref{fig:ls} is based on a exhaustive $\gamma\gamma$-coincidence analysis described below. The fitting function employed to obtain the $\gamma$-ray intensities and their energies has, as a fixed parameter, the energy-dependent peak-width calculated from the systematics of known transitions of other reaction channels measured in the same experiment.

\begin{figure}
\includegraphics[width=.50\textwidth]{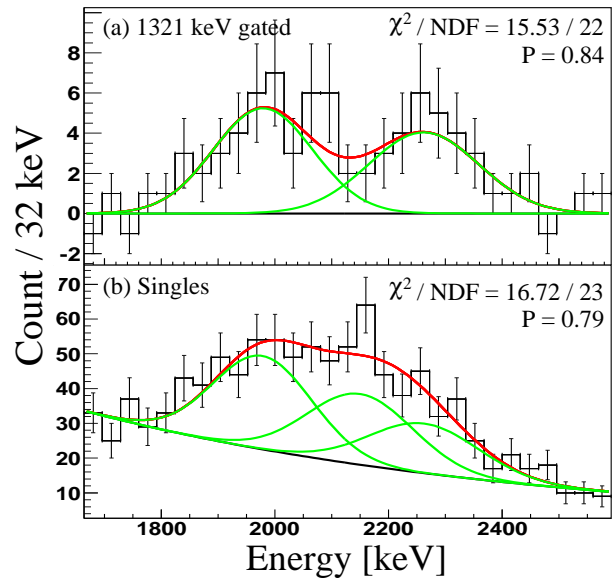}
\caption{\label{fig:fit} (Color online) High energy part of the spectra obtained (a) from the $\gamma\gamma$ matrix gated by the 1321~keV transition and (b) singles $\gamma$-ray. The lines shows the result of the fit with two (a) and three (b) Gaussian functions, respectively. The p-value of the $\chi^2$ goodness-of-fit test is represented by the letter P.}
\end{figure}

The spectrum obtained from the $\gamma\gamma$-matrix gated on the 2$^+_1\rightarrow$0$^+_1$ 1321~keV $\gamma$ transition is shown in Fig~\ref{fig:energy}(b). Two well separated $\gamma$ rays at high excitation energy are observed. A fit to the high~-~energy part of this spectrum with a two Gaussian functions yields 1979 and 2262~keV for the two $\gamma$-rays, respectively (Fig.~\ref{fig:fit} (a)). The p-value from the $\chi^2$ goodness-of-fit test is 0.84.
It was not possible to fit the broad high energy distribution visible in the singles spectrum with two Gaussian functions if the 1979 and 2262~keV centroids are considered as fix parameters (p-value~$<$~0.07). Similarly, if the high energy part of the singles spectrum is fitted first with two Gaussian functions and then the energies obtained are used as a fixed parameter in the fitting procedure of the spectrum gated on the 1321~keV, the p-value of the latter fit yields 0.3. The energies of the two $\gamma$ rays obtained by the two methods are thus not consistent.
Therefore, this analysis suggests that the high energy and broad $\gamma$-ray distribution visible in the single spectrum is actually composed of three different $\gamma$-rays.  
The fit of this distribution using three Gaussian functions, in which the centroids and widths of the 1979 and 2262 keV transitions observed in Fig.~\ref{fig:energy}(b) are fixed, yields 2156~keV as the energy of the third $\gamma$-ray (Fig.~\ref{fig:fit}(b)). These three transitions are visible with a significance level of more than 4~$\sigma$. The 2156~keV $\gamma$-ray is not in coincidence with the 1321~keV transition. Therefore, it is proposed that it decays directly from a level at 2156~keV to the ground state of $^{44}$S. The observation of a 1979~keV $\gamma$ ray connected to the 1321~keV 2$^+_1$ state establishes a level at about 3300~keV excitation energy (see below).
The structure in the $\gamma$-ray spectrum between 950 and 1050~keV is a doublet composed of two different transitions at 977 and 1006~keV [Fig~\ref{fig:energy}(c)]. The sum of the 977 and 1006~keV $\gamma$ energies leads to 1983(34)~keV which, within the experimental uncertainties, agrees with the energy of the 1979(19)~keV $\gamma$ ray. Therefore, these three transitions most likely belong to two parallel branches that de-excites from a new state. This state decays by a direct 1979 keV $\gamma$-ray and a cascade of 977 and 1006~keV transitions. The energy spacing between the proposed level and the 2$^+_1$ state at 1321~keV can be estimated using the weighted mean of the energies of the two decay branches and yields 3301(18)~keV. 

The 977~keV $\gamma$-ray is placed immediately above the 2$^+_1$ state based on the fact that its energy matches the one of 949(5)~keV reported in Ref.~\cite{santiago}. The 1006~keV transition is placed above the 977~keV and this $\gamma $-ray was not observed in Ref.~\cite{santiago}. This establishes the level at 2298(25)~keV corresponding that of 2268(5)~keV from Ref.~\cite{santiago}. 
The spectra shown in Fig~\ref{fig:energy}(b) and (c) reveal the possible existence of a new $\gamma$ ray at 1198~keV. This transition is likely in coincidence with all other $\gamma$ rays with the exception of the 2262 and 2156~keV peaks. It is placed above the 1979~keV $\gamma$ line depopulating a state at 4499~keV excitation energy with spin and parity Y$^+$ (see below).

\begin{table*}
\caption{\label{tab:inten} Experimental energies, tentative spin and parity assignments, relative intensities and transition energies for the excited states in $^{44}$S. The direct feeding to the measured levels have been normalized to the
 number of detected $^{44}$S in SPEG and was obtained considering the 2262~keV transition feeding the 1321~keV (D$_{A}$) or the (2$^+_3$) (D$_{B}$) state (see text). The width of the $\gamma$-ray ($\sigma$) used in the fitting function is shown in the last column.}
\begin{minipage}{10.3cm}
 \begin{tabular}{cccccccccccccc}
\hline
\hline
E$_i$ [keV] &I${_i^\pi}$ $\rightarrow$ I${_f^\pi}$&E$_\gamma$ [keV]& I$_\gamma[\%]$&D$_A$[\%]&D$_B$[\%]&$\sigma$[keV]&&\\
\hline
1321(10)&2$^+_1\rightarrow$0$^+_1$&1321(10)&100(8)&14(6)&20(7)&69(1)&\\
2156(49)&(2$^+_2$)$\rightarrow$0$^+_1$&2156(49)&17(6)&12(4)&12(4)&99(1)&\\
2298(25)&(2$^+_3$)$\rightarrow$2$^+_1$&977(23)\footnote[1]{Extracted from the $\gamma\gamma$ matrix gating on the 950~-~1050~keV distribution [Fig~\ref{fig:energy}(c)].}&48(6)&25(4)&11(4)&57(1)&\\
3301(18)&(2$^+_6$,1$^+_2$)$\rightarrow$(2$^+_3$)&1006(25)$^b$&12(3)&12(2)&13(2)&58(1)&\\
&(2$^+_6$,1$^+_2$)$\rightarrow$(2$^+_1$)&1979(19)\footnote[2]{Extracted from the $\gamma\gamma$ matrix gating on the 1321~keV transition [Fig~\ref{fig:energy}(b)].}&24(5)&&&93(1)&\\
3583(39)\footnote[3]{State depopulated by the 2262~keV transition feeding the 2$^+_1$ level.};4560(45)\footnote[4]{State depopulated by the 2262~keV transition feeding the (2$^+_3$) level. }&(X$^+$)$\rightarrow$2$^+_1$;(2$^+_3$)&2262(38)$^c$&21(5)&15(3)&15(3)&103(1)&\\
4499(37)&(Y$^+$)$\rightarrow$(2$^+_6$,1$^+_2$)&1198(25)$^b$&18(3)&13(2)&13(2)&65(1)&\\
\hline
\hline
\end{tabular}
 \end{minipage}
\end{table*}

From the present data, it is not possible to place the 2262~keV $\gamma$ ray in the level scheme. The coincidence relation with the 1321~keV line indicates that it must decay from a state above the 2$^+_1$ level, labeled X$^+$ in Table~\ref{tab:inten}.
The broad structure observed around 200-300~keV in the spectrum of singles (Fig.\ref{fig:energy}(a)) is not associated with $^{44}$S. This structure corresponds to a peak at constant energy in the non Doppler-corrected spectra and, consequently, must originate from a stopped source. As shown in  Fig.~\ref{fig:ls}, the present level scheme agrees with the one of Ref.~\cite{santiago} for the first three excited states. In addition, two new levels are proposed at higher energy in the present work that were not observed in Ref.~\cite{santiago} and highlights the different selectivity of the complementary reactions used in the two experiments.

The direct feeding to the levels normalized to the total number of $^{44}$S detected in SPEG have been
 extracted (Tab.~\ref{tab:inten}). They are based on the measured $\gamma$-ray intensities and calculated branching ratios. As the precise location of the 2262~keV $\gamma$-ray in the level scheme could not be determined unambiguously, the direct population to states in $^{44}$S were deduced considering two assumptions: (D$_A$) in which the 2262~keV transition feeds the 2$^+_1$ level or (D$_B$) in which the 2262~keV transition feeds the 2298~keV state. It is interesting to note that the direct feeding of all the states populated in $^{44}$S is nearly the same (between 10 and 20\%). This feature is a direct consequence of the large configuration mixing between levels in $^{44}$S. In contrast, the fact that none of the newly observed transitions is in coincidence with the 2156~keV $\gamma$ ray indicates a different configuration for the 2156~keV state.

\section{Discussion}

\begin{figure*}
\includegraphics[width=1.\textwidth,height=100mm]{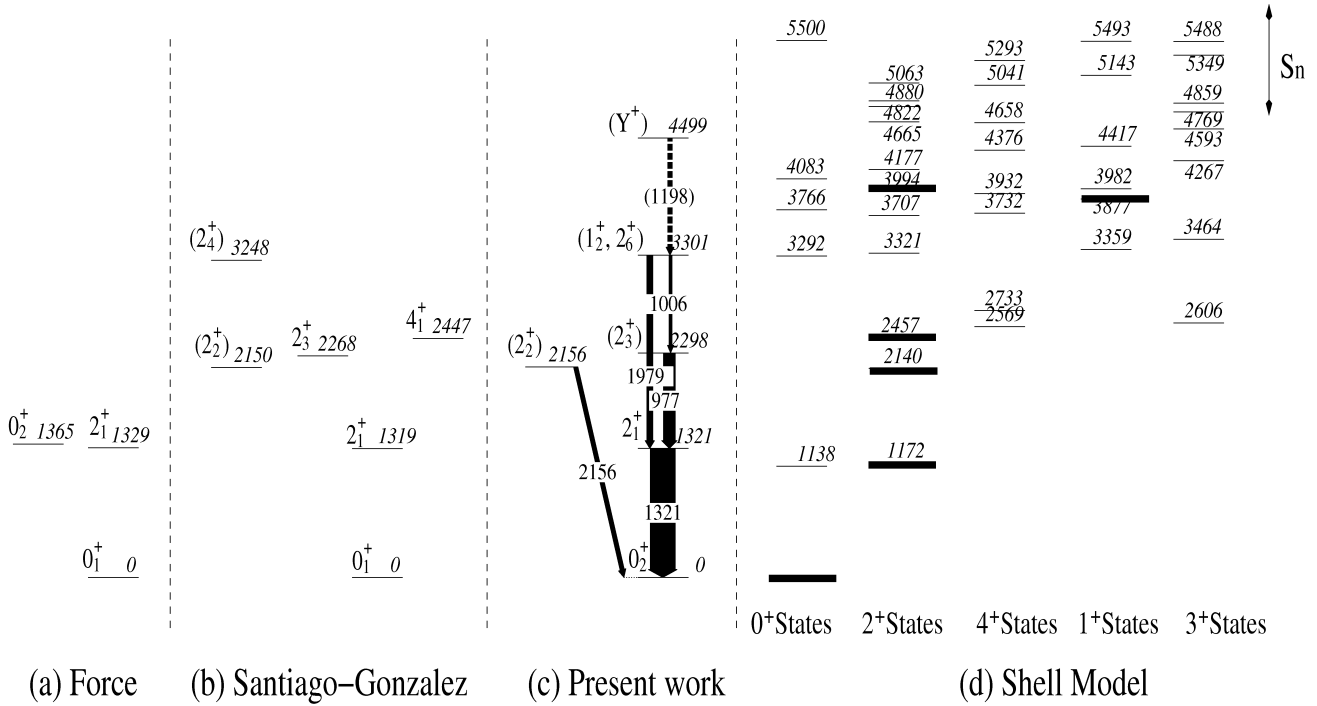}
\caption{\label{fig:ls}Experimental level scheme of $^{44}$S obtained from (a) Force:~Ref.~\cite{force}, (b) Santiago-Gonzalez:~Ref.~\cite{santiago} and (c) in this work. The width of the arrows in (c) indicates the relative $\gamma$-ray intensities normalized to the 1321~keV transition. The energy uncertainties on the $\gamma$-rays and levels are quoted in Table~\ref{tab:inten}. (d) Shell model calculations performed using the SDPF-U interaction. The arrow length in the right hand side of the figure shows the error on the neutron separation energy (S$_n$~=~5220(440)~keV~\cite{Sn}). The levels observed in this data are represented with thicker widths line.}
\end{figure*}

The tentative spin and parity assignment to the observed levels in $^{44}$S is based on the comparison between the calculated and experimental (i) excitation energies, (ii) branching ratios and (iii) direct population of the states following the one proton knock-out reaction. 
Shell model calculations (Fig.~\ref{fig:ls}) using the SDPF-U interaction~\cite{SDPF},
 known to reproduce the spectroscopic information in the region, were performed. The full model space was considered,
 comprising the proton $\pi$(d$_{5/2}$, d$_{3/2}$, s$_{1/2}$) and neutron $\nu$(f$_{7/2}$, p$_{3/2}$, p$_{1/2}$,
 f$_{5/2}$) orbitals. Calculations were performed with the ANTOINE code~\cite{antoine1,antoine2}. $E2$ reduced transition probabilities were calculated using polarization charges of 0.35~$e$ for both protons
 and neutrons. Effective magnetic moments for the $M1$ transition rate calculations are taken from
 Ref.~\cite{SM} and were derived from systematic studies in the $sd$ shell. They correspond to the bare values quenched by 0.75 for the spin operator while the orbital operator is set 1.1 and -0.1 for proton ($g_p$) and neutron ($g_n$) gyromagnetic moments, respectively. These values ensure a very good agreement with the measured g-factors in $^{44}$Cl~\cite{rydt} 
and $^{43}$S~\cite{43S}. 

\begin{figure}
\includegraphics[width=.550\textwidth,height=60mm]{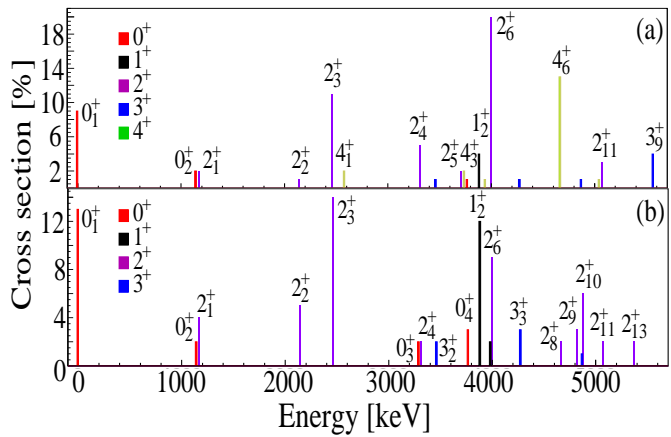}
\caption{\label{fig:SF} (Color online) Theoretical partial cross sections to the levels in $^{44}$S normalized to the strength below the neutron separation energy S$_n$~=~5220(440)~keV assuming 3/2$^+$ (a) and  1/2$^+$ (b) spin and parity assignment for the ground state of $^{45}$Cl.}
\end{figure}

Theoretical direct population of the levels in $^{44}$S has been obtained combining the eikonal-model calculations~\cite{tostevin} for the single particle cross sections with the shell model spectroscopic factors (SF) following the prescription of Ref.~\cite{gade08}. The ground state spin and parity of $^{45}$Cl are not firmly known experimentally~\cite{sorlin2,gade2}. Shell model calculations predict an I$^\pi$=1/2$^+$ assignment although the 3/2$^{+}$ level is calculated to be at only 132~keV excitation energy. Therefore, both spins (1/2 or 3/2) were considered in the SF calculations (Fig.~\ref{fig:SF}). The range of possible states populated in $^{44}$S from the proton knock-out reaction of $^{45}$Cl are from 0$^{+}$~to~3$^{+}$ (0$^{+}$~to~4$^{+}$) assuming 1/2$^+$ (3/2$^+$) for the ground state of $^{45}$Cl. Therefore, all calculations were restricted to these spin states~(Fig.~\ref{fig:ls}). 

The calculated 2$^+_1$ level at 1172~keV excitation energy corresponds to the 1321~keV experimental state previously reported in Ref.~\cite{glas}. 
Between 2000 and 2800~keV five levels were calculated. Among them, only the 2$^+_2$ state at 2140~keV excitation energy has a spherical configuration with intrinsic quadrupole moment -0.3~efm$^2$~\cite{force}. It is predicted to decay by 83\% and 16\% to the 0$^+_1$ and 2$^+_1$ levels, respectively. Its decay to the 0$^+_2$ level, although favored in terms of reduced transition probability (B(E2:~2$_2^+\rightarrow$0$^+_2$)$\simeq$~3.2~B(E2:~2$_2^+\rightarrow$0$^+_1$)), amounts to only 1\%. This is due to the energy dependence of the transition probabilities. Although, having comparable configurations, the 2$^+_2$ and 0$^+_2$ are thus only weakly connected to each other. 
The experimental analogue of the 2140~keV state is the 2156~keV level and the non observation of the 835~keV $\gamma$ ray associated with the 2$^+_2\rightarrow$2$^+_1$ decay is in agreement with experimental results considering the intensity of the 2156~keV $\gamma$-ray. The 2156~keV state is tentatively assigned to be the (2$^+_2$) level of spherical character based on the fact that if this state had had a more mixed configuration, higher lying states would have decayed through it and been detected in this experiment. 
As shown in Fig.~\ref{fig:SF}, the calculated direct population to the 2$^+_2$ state is 5\% and 1\% assuming 1/2$^+$ and 3/2$^+$ spin and parity assignment for the ground state of $^{45}$Cl, respectively. The theoretical direct feeding to the 4$^+_1$ state (calculated to be 2569~keV and observed at 2447(5)~keV in Ref.~\cite{santiago}) is 0\% in the case of a 1/2$^+$ and 2\% in the case of a 3/2$^+$ spin and parity of the ground state of $^{45}$Cl, respectively. Both, the experimental feeding of 12~\% to the 2$^+_2$ and the non observed feeding to the 4$^+_1$ state rather favor a I$^\pi$~=~1/2$^+$ ground state for $^{45}$Cl.
The other four calculated states between 2000 and 2800~keV (i.e. 2$^+_3$, 3$^+_1$, 4$^+_1$, 4$^+_2$) are predicted to decay mostly to the 2$^+_1$ level ($>$90\%); only the 2$^+_3$ state is expected to be strongly populated in the reaction (Fig.~\ref{fig:SF}), independent of the I$^\pi$ value of the ground state of $^{45}$Cl. For this reason, an I$^\pi$~=~(2$^+_3$) spin and parity is assigned to the 2298~keV level. Above 3000~keV there is a large level density is predicted. A study of the decay modes of each state was performed considering the theoretical reduced transition probabilities and the experimental excitation energies (when available), otherwise they were taken from theory. In the latter case, the branching ratio of each level was investigated considering a variation of 500~keV of the theoretical excitation energies. Only the 1$^+_2$ and 2$^+_6$ states decay by competing and parallel branches to the 2$^+_3$ and 2$^+_1$ levels in agreement with the experimental data.!
  Therefore, both states are possible candidates to spin and parity assignment for the 3301~keV level. In addition, the theoretical cross section calculations further support the I$^\pi$~=~(1$^+_2$, 2$^+_6$) assignment (Fig.~\ref{fig:SF}). Due to the large number of calculated levels around 4000~keV, it was not possible to find a theoretical counterpart to the experimental 4499~keV level, therefore no spin assignment can be given.

\section{Summary}
The $^{44}$S nucleus has been investigated using in-beam $\gamma$-ray spectroscopy following the one proton knock-out reaction from $^{45}$Cl. The large efficiency of the BaF$_2$ detectors of the Ch\^{a}teau de Cristal array, together with the particle selection provided by the SPEG spectrometer, allowed the construction of the level scheme up to high excitation energy. Analysis of $\gamma\gamma$-coincidences relations have been realized in this nucleus and the experimental direct feedings of the states were extracted. State-of-the-art nuclear shell model calculations have been performed for comparison to experimental energies, deduced spin assignments and decay branches of the states, and estimate their direct feeding from the one proton knock-out reaction. The energies and tentative spin assignments of the three excited states (2$^+_1$, 2$^+_2$ and 2$^+_3$) determined in Refs.~\cite{force,santiago} are confirmed. The 2$^+_2$ level at 2156(49)~keV is found to have a specifi!
 c configuration, and is likely spherical. This is based on the weak feeding observed from other excited states, its feeding intensity from the (-1p) knock-out reaction and its decay pattern.
Two new excited states have been proposed at higher excitation energy with the 3301~keV level tentatively assigned to have I$^\pi$~=~(1$^+_2$, 2$^+_6$), while no spin and parity could be determined for the level at 4499~keV excitation energy. 
The information from  the present data, and the one reported in Ref.~\cite{force,santiago}, highlights the large density of states at low excitation energy in the semi-magic $^{44}$S nucleus. This confirms the complexity of its nuclear structure that combines both shape and configuration coexistence.

\begin{acknowledgments}
This work benefits from discussions with G.~F.~Grinyer. 
The authors are thankful to the
GANIL and LPC staffs. We thank support by BMBF
No. 06BN109, GA of Czech Republic No. 202/040791,
MEC-DGI-(Spain) No. BFM2003-1153, and by the EC
through the Eurons Contract No. RII3-CT-3/2004-506065
and OTKA No.~K68801 and Bolyai J\'anos Foundation. R. B., A. B.,
M. L., S. I., F. N. and R. C., X. L. acknowledge the
IN2P3/CNRS and EPSRC support.
\end{acknowledgments}



\end{document}